\documentclass[10pt,twocolumn]{article}

\usepackage{epsfig} 
\usepackage{graphicx}
\usepackage{grffile}
\usepackage{mathtools}
\usepackage{amsmath}
\usepackage[intoc]{nomencl}
\usepackage[utf8]{inputenc}

\title{\textbf{ An Overview of Computational Fluid Structure Interaction: Methods and Applications }}
\author{
    \textbf{Sumant R. Morab} \\
    Doctoral Student\\
    Department of Mechanical Engineering\\
    Indian Institute of Technology, Bombay\\
    Email: sumantmorab1996@gmail.com\\
  \and
    \textbf{Atul Sharma}\thanks{Address all correspondence related to this author.}\\
    Professor\\
    Department of Mechanical Engineering\\
    Indian Institute of Technology, Bombay\\
    Email: atulsharma.iitb@gmail.com
}
\date{}
\makenomenclature

\begin{document}
	
	\maketitle    
	
		{\it Over the past few decades, there has been a rapid improvement in computational power as well as techniques to simulate the real world phenomenon which has enabled us to understand the physics and develop new systems which outperform the existing ones. In the domain of multi-physics problems, fluid and structure interactions have been studied and various numerical methods are introduced to solve them. An extensive review of most commonly employed numerical techniques to solve fluid structure interaction(FSI) problems has been done in this article. It also becomes utmost important to understand the applicability of each method and hence a critical reasoning for usage of different methods has been provided. Application domains are presented which range from energy harvesting processes to simulation of human vocal cords and movement of bolus(food) inside esophagus. Suitable numerical methods applied for each application have also been discussed. Numerical methods developed so far are classified in particular groups based on the discretization, the manner in which coupling is performed and on the basis of meshing(division of entire domain into small blocks inside which variation of a variable is approximated). Challenges and instabilities posed by presently available numerical methods are discussed and potential applications where there is a possibility of errors due to these methods have been listed. A brief set of application areas which have not been explored through the lens of fluid structure interaction also have been discussed at the end of this article.
			\\
			Keywords: Multi-physics problems, Fluid Structure Interaction(FSI), Bolus transport, Discretization.\\\\		
		}
		\textbf{Nomenclature\\}
		\begin{nomenclature}
		{A~~}{Displacement in solid\\}
		{e~~}{Shear strain in fluid\\}
		{E~~}{Youngs Modulus of solid\\}
		{f~~}{External forces\\}
		{G~~}{Bulk Modulus of solid\\}
		{n~~}{Normal direction to surface\\}
		{p~~}{Hydrodynamic pressure\\}
		{t~~}{time\\}
		{v~~}{Velocity\\}
		{x~~}{co-ordinates\\}
		
		{superscripts\\} {}
		{f~~}{fluid domain\\}
		{s~~}{solid domain\\}

		subscripts\\
		{i,j~~}{directions in tensor form\\}
		
		{Greek Symbols\\} {}
		{$\delta~~$}{Kronecker delta\\}
		{$\epsilon~~$}{lateral strain inn solid\\}
		{$\lambda~~$}{Lame Constant\\}
		{$\mu~~$}{Dynammic viscosity\\}
		{$\nu~~$}{Poissons ratio\\}
		{$\Omega~~$}{closed surface/volume\\}
		{$\rho~~$}{denisty\\}
		{$\tau~~$}{Shear stress\\}
		{$\Gamma~~$}{Interface\\}
		{$\sigma~~$}{Hydrodynamic stress}
	\end{nomenclature}

	\printnomenclature
	\section{Introduction}
	
	Major research in the domain of Energy Harvesting, Transportation and Medical science involves multi-physics and interactions between a fluid and a solid. Modern research in disease diagnosis and treatment has helped us to take a
	strong forward leap in healthcare and this medical research owes a part to fluid structure
	interaction. Interaction between a fluid and a solid occur on day to day experiences like
	movement of leaves due to wind, movement of flag in air, production of sound etc. and in
	most complex engineering applications like flutter of airfoil and energy harvesting.
	\\
	\\
	Fluid Structure Interaction (FSI) is a class of multi-physics problem where
	dependence between fluid and structural part plays an important role. The deformation and
	displacement of a structure is affected by the forces applied by the fluid flowing past through
	it and change in geometry or position of solid inturn has an effect on fluid flow pattern. Thus,
	there is a two way coupling that exists between fluid and solid and capturing this through
	computational techniques can have humongous applications in engineering design and
	medical field. Various examples of FSI found in our nature include locomotion of aquatic
	animals, chirping and flying of birds, dispersal of pollen and seeds by wind, circulation of
	blood through the entire body, beating of heart etc. which has inspired humans to
	mimic them and build efficient machines. Engineering applications of FSI include
	development of energy harvesting devices, designing underwater vehicles to decrease drag
	and improve performance, design of turbine blades for various applications, design of high
	rise buildings to avoid vibrations etc. moreover in medical field, we find applications in
	diagnosis of various cardiovascular diseases (CVD) like atherosclerosis, arteriosclerosis,
	aneurysms, atherogenesis and prediction of future growth prospects. In recent years, FSI has
	been used to simulate sound production from vocal cords and determine the physics behind
	bolus transport in esophagus. Also, there has been growing use of FSI for surgical planning
	and treatment in various sectors. The above mentioned applications have been discussed in
	detail in later sections of this report. Thus, it becomes very important to develop predictive
	FSI techniques that provides solutions to the major concerns faced by modern era.
	\\
	\\
	For independent problems on fluid or solid, there have been some
	analytical results \cite{drazin2006navier} which help us to
	understand the physics. In case of coupling between fluid and solid, there have been rarely
	any analytical solutions except by Womersley \cite{womersley1955method}. Another possibility
	of experimentation has been carried out for some applications like vortex induced vibration
	\cite{feng1968measurement} but when it comes to medical applications, invivo experimentation on humans seems highly impossible. Thus, in such cases medical
	images obtained from computed tomography (CT) or magnetic resonance angiography(MRA)
	have been used to construct three dimensional geometry and perform computational
	experiments. From the above argument we can infer that development of computational
	techniques for FSI problems becomes very crucial and taking the advantage of huge contribution by scientific community towards the development of various
	numerical schemes which can efficiently handle the complexities involved in geometry,
	material and fluid properties becomes crucial. The major challenge for FSI problems is the coupling between
	structure and fluid beside ensuring that boundary conditions are implemented accurately at the
	fluid and solid interface. There have been different set of models developed to tackle this
	issue and each one of them along with their applicability for certain kinds of problems has
	been briefed in the next section.	
	
	\section{Mathematical Formulation}
	
\begin{figure}[t]
		\centerline{\psfig{figure=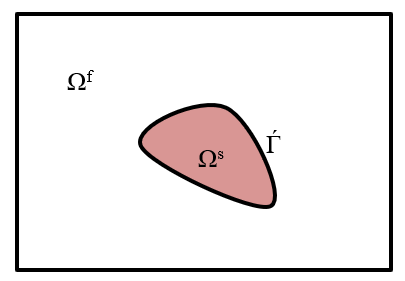,width=3.25in}}
		\caption{Computational Domain}
		\label{figure_domain} 
	\end{figure}

	Consider the whole domain of interest to be split into Solid ($\Omega$\textsuperscript{s}) and fluid ($\Omega$\textsuperscript{f}) with interface between them($\Gamma$) as shown in Fig.~\ref{figure_domain}. The boundaries of fluid not in contact with fluid are assumed to be rigid and thus a standard boundary condition may be applied on them. The representation of the symbols presented below are provided in the nomenclature.
	
	\subsection{Fluid Dynamics}
	
	It is assumed that the fluid is in-compressible and possesses a Newtonian rheology. The mass conservation associated with each cell leads to  Continuity given by
	
	\begin{equation}
	\frac{\partial \rho v_{i}^{f}}{\partial {{x}_{i}}}=0
	\label{eq_mass}
	\end{equation}
	
	Momentum conservation at Control volume leads to
	
	\begin{equation}
	\rho \frac{Dv_{i}^{f}}{Dt}=\frac{\partial \sigma _{ij}^{f}}{\partial {{x}_{j}}}-f_{i}^{f}
	\label{eq_mom}
	\end{equation}
	
	Where
	\begin{equation*}
	\begin{multlined}
	\frac{D{{v}_{i}}}{Dt}=\frac{\partial {{v}_{i}}}{\partial t}+{{c}_{j}}\frac{\partial {{v}_{i}}}{\partial {{x}_{j}}} \\ 
	{{\sigma }_{ij}}=-p{{\delta }_{ij}}+{{\tau }_{ij}} \\ 
	{{\tau }_{ij}}=2\mu ({{e}_{ij}}-{{\delta }_{ij}}{{e}_{kk}}/3) \\ 
	{{e}_{ij}}=\frac{\partial {{v}_{j}}}{\partial {{x}_{i}}}+\frac{\partial {{v}_{i}}}{\partial {{x}_{j}}} \\    
	\end{multlined}    
	\label{eq_mom2}
	\end{equation*}
	
	\subsection{Structure Dynamics}
	
	Consider the solid strcuture ($\Omega$\textsuperscript{s}) to be elastic, homogenous and isotropic. According to principle of virtual work , control volumes of solid have to satisfy
	
	\begin{equation}
	{{\rho }^{s}}\frac{Dv_{i}^{s}}{Dt}=\frac{\partial \sigma _{ij}^{s}}{\partial {{x}_{j}}}-f_{i}^{s}
	\label{eq_solid}
	\end{equation}
	
	Where
	\begin{equation*}
	\begin{gathered}
	\sigma _{ij}^{s}=\lambda {{\delta }_{ij}}{{\varepsilon }_{ll}}+2G{{\varepsilon }_{ll}} \\ 
	{{\varepsilon }_{ij}}=\frac{1}{2}\left( \frac{\partial {{u}_{i}}}{\partial {{x}_{j}}}+\frac{\partial {{u}_{j}}}{\partial {{x}_{i}}} \right) \\ 
	G=\frac{E}{2(1+\nu )} \\
	\lambda =\frac{E\nu }{2(1+\nu )(1-2\nu )}
	\end{gathered}
	\end{equation*}
	
	\subsection{Coupling equations for Interface}
	There are mainly two conditions which have to be satisfied at the interface of any solid and fluid.
	\begin{enumerate}
		\item Velocity of fluid boundary at interface must be same as the velocity of solid boundary(velocity due to its displacement). This is generally termed as kinematic boundary condition or no slip condition.
		\item The stress exerted by fluid at the interface should be same as the external force which acts on solid. this is termed as Dynamic Boundary Condition or stress continuity constraint.
	\end{enumerate}
	No Slip and stress continuity is applied at the interface($\Gamma$) as shown.
	
	\begin{equation*}
	\begin{multlined}
	v_{i}^{s}=v_{i}^{f} \\ 
	\sigma _{ij}^{s}{{n}_{i}}=\sigma _{ij}^{f}{{n}_{i}} \\
	\end{multlined}
	\label{eq_interface}
	\end{equation*}
	
	\section{Computational FSI Methods}
	There have been various computational methods to solve equations governing the fluid dynamics \cite{sharma2016introduction} (usually the Navier-Stokes equation) and structure dynamics separately. They have been broadly called as Computational Fluid Dynamics (CFD) and Computational Structure Dynamics (CSD). Coupling between them started to pick up eventually when people started to think about the reason behind unfortunate events like the collapse of the Tacoma Narrows bridge, Vibration of deep water risers in oil rigs etc. and eventually computational techniques started getting developed. Major classification of numerical methods is based on the way coupling between fluid and solid is introduced, the way grid is generated (i.e, whether it changes with fluid/solid domain) and the way in which discretization has been performed to obtain the final algebraic equations to be solved. These classifications are discussed in detail below.
	
	\subsection{Monolithic and Partitioned Approach}
	
\begin{figure}[t]
		\centerline{\psfig{figure=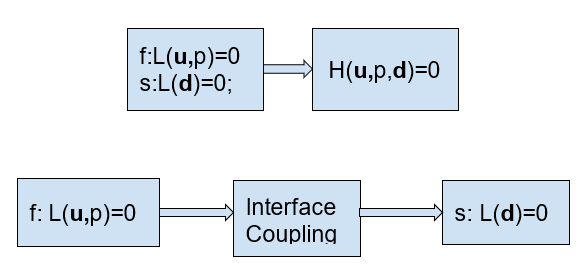,width=3.25in}}
		\caption{Monolithic and Partitioned Approach}
		\label{figure_mono_parti} 
	\end{figure}

	This classification is made on the basis of how the mathematical framework has been designed to solve for fluid and solid domains. In case of Monolithic approach a unified mathematical governing equations are defined for both fluid and solid and the whole domain is solved as one entity. The main advantage behind using this method is that the interfacial conditions are inherently present in the formulated mathematical framework and a single discretization scheme can be applied. It has also been observed \cite{hubner2004monolithic} that a better accuracy is observed in case of well defined multi-physics problems. One of the main issues with this approach is that the code development becomes heavily problem-specific and a sense of generality is lost. Also, some kind of expertise is required in the domain specific area to develop such an algorithm. Fig. \ref{figure_mono_parti} shows the basic flow in these two approaches.
	\\
	\\
	In case of partitioned approach, solid and fluid governing equations are solved separately in their respective domains and coupling between them is introduced by applying suitable interface conditions. Discretisation scheme and meshing can be completely different for both the domains but a sense of match must exist at interface location. If meshes are not in a matching condition at the interface, a ghost patch \cite{hou2000analytical} is usually introduced wherein communication is transmitted from solid to fluid or vice versa through this ghost patch. This ghost patch is developed in such a way that it matches solid and fluid domains on their respective sides. This method was developed in the beginning stages of computational FSI development and is still being applied and developed since it has an inherent advantage of using robust fluid and solid solution algorithms which have been developed over the past few decades. These methods are prone to errors especially at the interfaces due to mismatch of discretisation and mesh. Another issue with this method is that the computational cost is high compared to that of Monolithic approach since interface conditions needs to be applied every time and remeshing of the whole domain has to be done either for every time step or when there is a large displacement which could possibly disturb the skewness and other properties of the mesh.
	
	\subsection{Conforming and Non-Conforming Mesh}
\begin{figure}[t]
		\centerline{\psfig{figure=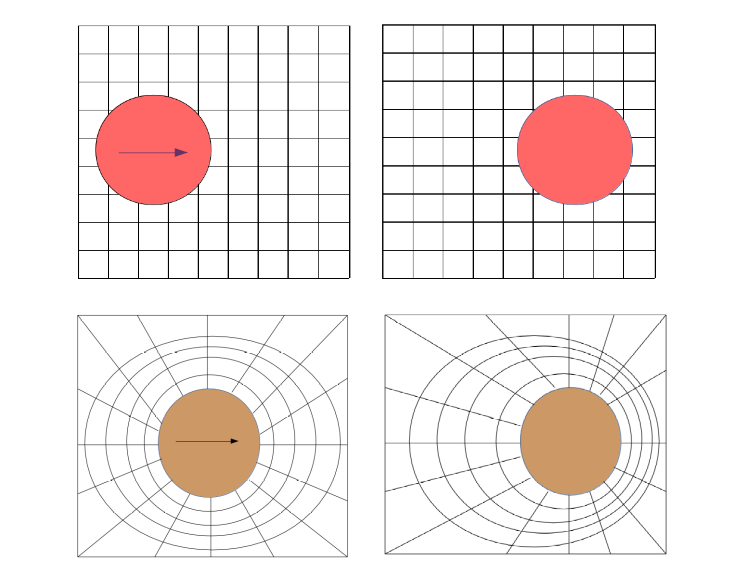,width=3.25in}}
		\caption{Non-Conforming (top) and Conforming (Bottom) mesh}
		\label{figure_conform_nonconform} 
	\end{figure}

	A classification based on the type of mesh generated to solve fluid solid interactions leads us to conforming and non-conforming type mesh (Fig. \ref{figure_conform_nonconform}). In case of conforming type system, the interface between fluid and solid is considered to be a physical boundary. In most of the cases, fluid particles are tracked through Eulerian approach whereas solid movement is tracked via Lagrangian formulation. Boundary conditions are applied at the interface for both fluid and solid governing equations and thus the communication between both the domains happens through these boundary conditions. Fluid and solid domains can utilise various mesh types but needs to be updated at every/regular time-steps. For example, if a body contracts the mesh needs to be updated to either remove the solid cells outside contracted body or to contract the existing mesh. This task becomes cumbersome when there are a large number of elements in mesh and deformation is high. Tracking of the boundary is also essential while going to the next time step. Methods under this category include Arbitrary Lagrangian Eulerian (ALE)(\cite{hirt1974arbitrary}, \cite{kennedy1982theory}), Deformed Spatial Domain/Stabilised space time (DSD/SST) \cite{tezduyar1992new} etc. which will discussed in the upcoming topics.
	\\
	\\
	In case of non-conforming mesh methods, the interface conditions are not treated as physical boundary conditions for fluid and solid domain. They are treated in governing equations itself so as to give the same effect of boundary conditions (no slip and specified traction). The major advantage with these methods is that though fluid and solid are discretized quite differently, a single mesh is enough to solve the problem of FSI. Thus, there won’t be any issue of re meshing in this type which would be very effective in reducing computational cost. Efficient mathematical formulation of governing equation for accurate interface conditions needs to be considered. Various methods have been developed under this category which include Immersed Boundary Method(IBM) \cite{peskin2002immersed}, Immersed Interface Method(IIM) \cite{li2001immersed}, Coupled-Momentum method \cite{figueroa2006coupled}, Direct Forcing method etc. which will be detailed in the next topic.
	
	\subsection{Based on Numerical discretisation and Interface coupling schemes}
	
	After the development of some initial Fluid-Structure-Interaction techniques based on a naive approaches, various discretisation techniques started getting developed for specific problems to improve accuracy. Some of these methods which are applicable for general FSI problems are discussed below.
	
	\subsubsection{Arbitrary Lagrangian Eulerian (ALE) methods}
	In the initial stages, this method was developed to obtain solutions for interfacial and free surface flows. Hirt et al. \cite{hirt1974arbitrary} brought ALE method which was applicable to all flow regimes. It used a Finite Difference discretisation in which nodal points could move with fluid or be held at fixed position or could be moved in a specific direction as specified by user. This led to the generation of word ‘Arbitrary Lagrangian Eulerian’. Based on the value of mesh velocity, the discretization became either fully Eulerian or fully Lagrangian or hybrid. The computations in a particular time step have been performed in three separate phases. These phases include explicit velocity computation, implicit calculation of pressure and updating the velocity and rezoning using mesh movement. The method was tested on several cases which included shock tube flow with moderate and very large sound speed and flow around a rectangular block. Stability analysis was also performed with this method.
	\\
	\\
	Kennedy et al. \cite{kennedy1982theory} developed a computational method to analyse the effect of pressurised bubble on a structure used in reactor vessels. They used Finite Element discretization on the fluid domain and for energy equation. After obtaining fluid solution, displacements of solid was calculated through traction force exerted by fluid. After this step, mesh velocity and co-ordinates were updated to perform remeshing. The numerical method was tested on a fluid inside a rigid and flexible walls. A brief algorithm followed by them is as follows:
	\\
	\begin{enumerate}
		\item 
		Calculate acceleration of fluid as
		\begin{equation}
		\frac{\partial \mathbf{v}}{\partial t}={{\mathbf{M}}^{-1}}\left( {{\mathbf{f}}^{ext}}-{{\mathbf{f}}^{\operatorname{int}}} \right)
		\label{eq_ASME1}
		\end{equation}
		
		\textbf{M} is found through Navier Stokes equation (2.2).
		\item
		Obtain fluid velocity \textbf{v}
		\begin{equation}
		{{\mathbf{v}}^{n+1/2}}={{\mathbf{v}}^{n-1/2}}+\Delta {{t}^{n}}\frac{\partial {{\mathbf{v}}^{n}}}{\partial t}
		\label{eq_ASME2}
		\end{equation}
		\item
		Set Grid Velocity \textbf{v} \textsuperscript{G} as some fraction of fluid particle velocity at interface obtained
		through structure solution.
		\item
		Update mesh point co-ordinates as
		\begin{equation*}
		\begin{multlined}
		\mathbf{x}=\mathbf{x}+\Delta {{t}^{n+1/2}}{{\mathbf{v}}^{G,n+1}} \\ 
		\Delta {{t}^{n}}=\frac{1}{2}\left( \Delta {{t}^{n+1/2}}+\Delta {{t}^{n-1/2}} \right) \\
		\end{multlined}
		\label{eq_ASME}
		\end{equation*}
		\item
		March to the next time step.
	\end{enumerate}
	
	Most commonly used discretization scheme for solids has been finite elements because of its robustness to handle material and geometric nonlinearities. There have been some attempts \cite{fallah2000comparison} to discretize solid through Finite Volumes (FVM) and compare the accuracy with that of Finite Element Method(FEM). The main motivation behind using such a discretization is to allow efficient and smooth coupling at the interface of solid and fluid (which is mostly discretized through Finite Volumes in an Eulerian frame). Tsui et al. \cite{tsui2013finite} adopted Finite Volume discretization and partitioned approach for both fluid and solid domains. Their method had a contrasting uniqueness where fluid variables were placed at grid centroids whereas solid variables were at the nodes of cell. They successfully validated results for flow past a cylindrical bluff body with plate fixed on the rear side.\\
	
	Hubner et al.\cite{hubner2004monolithic} developed a monolithic approach for FSI problems where-in simultaneous solution was achieved for fluid and solid. Remeshing was performed before marching to the next time step. Discretization was performed through space time finite elements \cite{hughes1988space}. The simultaneous solution was possible by formulatining fluid and solid discretized equation into a single equation. The solid was modelled with a complete Lagrangian fashion where displacement and Piola-Kirchhoff stresses were the variables to be solved for. Fluid domain was modelled in an Arbitrary-Lagrangian Eulerian representation where boundaries of fluid domain intersecting with solid were in Lagrangian mode. The method was validated by solving flow over a building roof, and through vortex excited elastic plate.
	\\\\
	Though these methods are found to be accurate and easier(as they use existing solvers of fluid and solid domain), remeshing has to be carried out at successive interval time which increases the computational cost by a huge margin. These methods encounter issues for interface matching due to which significant accuracy at interfaces is lost. When we deal with moving bodies (eg. object falling under gravity) or with large deformation problems (expansion of air inside balloon etc.,), mesh generated at different intervals may be of lesser order or lead to entanglement (when solid rotates). Thus, in these cases, ALE is found to be not very suitable.
	
	\subsubsection{Immersed Boundary Methods (IBM)}
	
\begin{figure}[t]
		\centerline{\psfig{figure=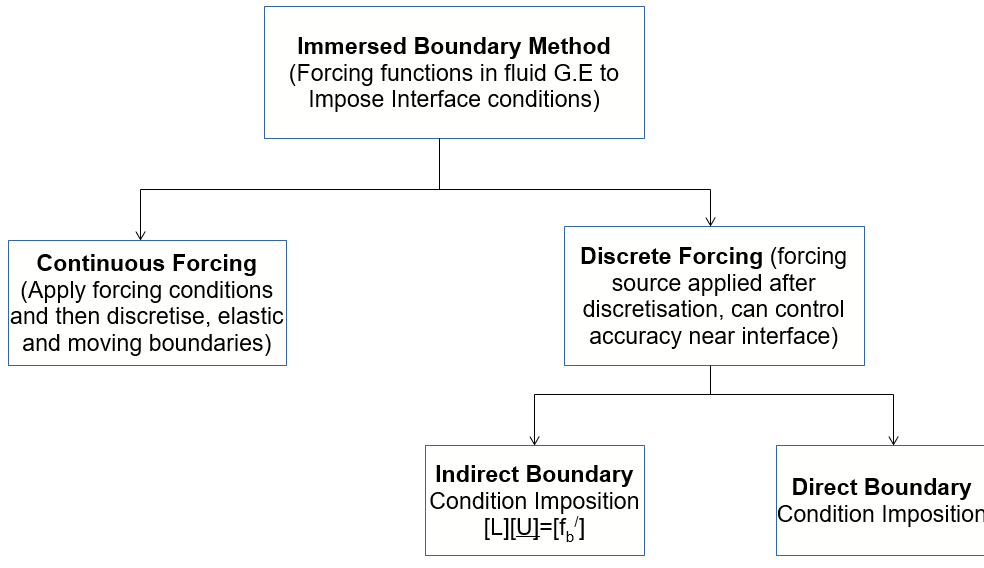,width=3.25in}}
		\caption{Flowchart of Immersed Boundary Method}
		\label{figure_IBM} 
	\end{figure}

	The IBM is a Non-Conforming methods that uses a Cartesian grid on whole domain and boundary is approximated either as a curve(for 2D flow) or a surface(3D flow). Structure is solved separately to update the coordinates of curve. Further, the governing equations in IBM are modified so as to incorporate the boundary conditions of no-slip by introducing source terms. This method was introduced by Peskin \cite{peskin1977numerical} where blood flow patterns in a model heart was analysed. The method used Eulerian frame for fluid (blood) that is discretized by finite differences and applied Chorin’s projection method for obtaining fluid solution. Solid was modelled in Lagrangian frame where forces generated were modelled through combination of springs.
	\\\\
	Immersed Boundary Method is broadly classified in two types based on the manner in which forcing term is applied (refer Fig. \ref{figure_IBM}). In the first method, which is popularly known as ‘Continuous Forcing’, force term developed through solid at the interface is applied on the fluid momentum equation (Eqn.~(\ref{eq_mom})) and then a suitable discretization is performed. The following equations may help to visualise the flow of solution scheme with this approach—
	\begin{equation}
	[L]({{v}_{i}})=\frac{\partial {{v}_{i}}}{\partial t}+{{v}_{j}}\frac{\partial {{v}_{i}}}{\partial {{x}_{j}}}+\frac{1}{\rho }\frac{\partial p}{\partial {{x}_{i}}}-\frac{\mu }{\rho }\left( \frac{{{\partial }^{2}}{{v}_{i}}}{\partial x_{j}^{2}} \right)
	\label{eq_ibm1}
	\end{equation}

	A forcing term is introduced as follows
	\begin{equation}
	[L]\{{{v}_{i}}\}=\{{{f}_{b}}\}
	\label{eq_ibm2}
	\end{equation}
	\\
	The forcing terms are interpolated from solid domain using dirac delta functions as follows (k referes to the discrete point on interface curve).
	\\
	\begin{equation}
	{{\mathbf{f}}_{m}}(\mathbf{x},t)=\sum\limits_{k}{{{\mathbf{F}}_{k}}(t)\delta (|\mathbf{x}-\mathbf{X}|)}
	\label{eq_ibm3}
	\end{equation}
	\\
	Where ‘x’ denotes co-ordinates in fluid domain(eulerian) and ‘X’ denotes co-ordinates in solid field(Lagrangian). Various algebraic expressions are presented for ‘$\delta$’ by Peskin \cite{peskin2002immersed}. Velocity at the Lagrangian points is obtained through
	\\
	\begin{equation}
	\frac{\partial {{\mathbf{X}}_{k}}}{\partial t}=\mathbf{v}({{\mathbf{X}}_{k}},t)
	\label{eq_ibm4}
	\end{equation}
	\\
	It was observed by Mittal and Iaccarino \cite{mittal2005immersed} that continuous forcing approach was suitable for immersed elastic bodies and moving boundaries immersed in a fluid but produced errors when simulating rigid bodies and bodies with complex physical boundaries as it softens the interface.
	\\\\
	Another class of IBM are the discrete forcing approach based models where a normal fluid governing expression without forcing terms are discretized first and then a modified forcing term is applied to discretized equations via dirac delta kernels (so that force applies only on boundary and near boundary nodes). This method further has classifications like indirect boundary condition imposition  where fluid velocity is predicted without forcing terms and then corrected through suitable usage of kernel functions for force and direct boundary condition imposition methods which prevent smoothing of boundaries. This indirect boundary condition imposition is specially used for boundary layer problems like flow through vented cylinders, pectoral fin locomotion analysis. The broad classifications of IBM have been provided below.
	
	\subsubsection{Immersed Interface Method (IIM)}
	After the introduction of Immersed Boundary methods, it was observed that Pressure and viscous stress were discontinuous at the interface and IBM regularized this discontinuity. The main reason behind this was the usage of discrete dirac delta functions for singular force term calculation at the interfaces. Thus, it was found to have first-order accuracy at interfaces \cite{li2001immersed}. Though sharp interface IBM was developed \cite{peskin2002immersed}, full second order accuracy was not possible. The initial Immersed Interface Method (IIM) was developed by Li \cite{pouderoux1997timing} but was restricted to simpler elliptic PDEs applied to phase change problems and stokes equations. Li and Lai \cite{lai2000immersed} developed IIM for Navier-Stokes equation which was able to achieve second-order accuracy at interface. The major modification with respect to IBM was that the jump conditions of variables calculated through Taylor series were applied at the interface instead of discrete delta functions. The derivation for jump theorem and calculation of jump are reported in Li and LeVeque \cite{leveque1997immersed}. The method was successfully applied on moving interface problems that involved phase changes.
	\\\\
	A recent study by Griffith and Patankar \cite{griffith2020immersed} showed that there were some discrepancies in the solution obtained for flow past cylinders with IBM. They stated that this problem of leakage of mass flux inside cylinder could be easily overcome by the usage of IIM. When a channel flow was tried to be solved through IBM \cite{kolahdouz2020immersed}, it was observed that velocity diffused outwards of channel walls. They introduced a novel formulation for solving Fluid-Structure Interaction problems through Immersed Interface Method by calculating jump conditions from projections. Performing simulations with discrete IB formulation reveals that results near the interface have lower order accuracy and thus produce erroneous results. In case of Vortex Induced Vibration problem, it was observed to have some internal flow which is not desirable. In case of channel flow, deviation from analytical results were observed in flow rate with the use of classical IB formulation. They have projected the jump conditions on the subspace containing interface and then derived the jump conditions. This adjustment has been done to avoid the discontinuity in the interface normal directions. Ray-Casting Algorithm has been used to find the intersection between finite difference stencil and the interface. They have used IBAMR software to incorporate their new methodology and carry-out simulations. Many fundamental flow problems like flow in 2D and 3D channel (straight and inclined), flow over stationary and rotating cylinder have been solved with the proposed as well as classical IB method. Better results (in terms of sharpening of boundary and avoiding internal flow) were observed with IIM (with projected jump conditions applied on all variables). It was observed that solution was globally second order accurate for velocity and traction variables whereas the jump conditions with least order were considered using discretisation. This was achieved due to the projection of jump conditions across the interface. Thekkethil and Sharma \cite{thekkethil2019level} developed Level-Set based IIM formulation to solve FSI problems involving flexible bodies. Boundary conditions at interface were applied by the Level Set(LS) function which was calculated based on the geometry of solid.
	
	\subsubsection{Coupled Momentum Method}
	Although there have been some Non-Conforming techniques like IBM and IIM which can cut the computational cost and make programming clearer, coupling accuracy especially while using different discretizations can be error prone at interfaces. A new method which is monolithic in nature and avoids errors at interfaces was developed by Figueroa et al. \cite{figueroa2006coupled} and is popularly called as coupled momentum method. Using the thin wall approximation, the lateral body of solid surface (Eqn.~(\ref{eq_cmm}) ) is projected onto the fluid surface using quantitative wall thickness. This process is approximated as
	\\
	\begin{equation}
	\int\limits_{{{\Omega }^{s}}}{(\ldots )dx=\zeta }\int\limits_{{{\Gamma }^{s}}}{(\ldots )ds}
	\label{eq_cmm}
	\end{equation}
	\\
	
	The Nominal governing equation for fluid is written in integral form of Finite Element discretisation approach. Traction boundary condition \textbf{t}\textsuperscript{f} is brought into the governing equations for fluid. It was assumed that surface traction \textbf{t}\textsuperscript{f}  acting on the lateral surface is equal and opposite to vessel wall surface traction i.e, \textbf{t}\textsuperscript{f}  = - \textbf{t}\textsuperscript{s} . This inturn was incorporated in body force term of solid governing equation and an expression for \textbf{t}\textsuperscript{f}  was obtained. This force was substituted in fluid-governing equation. Thus, a common governing equation for fluid and solid was thus obtained. 
	Simulations were performed on idealized model of common carotid artery and deformation model was compared with that of rigid model. Resistance, Impedence and Pressure type boundary conditions were used and displacement of artery wall along with flow rates were measured. Recently a verification study for this method was published by Filonova et al. \cite{filonova2019verification}. Some challenges faced by this method include incorporating spatially varying material properties and development of method for varying thickness models.
	
	
	\section{Computational FSI Applications}
	
	In this section, an attempt has been made to cover some important applications involving interactions between fluid and solid. The research gap which exists in individual application domains have been tried to be addressed so that it would pave way for working on those areas to provide solutions. The applications are presented below in separate subsections for VIV induced energy harvesting, nature inspired vehicles and bio-medical applications.
	
	\subsection{Vortex Induced Vibrations}
	
	After the incident of Tacoma Narrows Bridge in 1940, scientific community started to study the reason behind the collapse of reason. One theory suggested by community suggested the shedding frequency of swirls behind the bridge were actually matching the natural frequency of vibration of bridge because of which it started to resonate. The study of vortices and the effect it has on a structure thus began which led to the development of study related to interaction between fluid and solid. Study of VIV has been applied in various fields which include prevention of vibrations of vertical structures like chimney stack, offshore risers and harvesting energy (called vortex power).
	\\\\
	Vortex induced vibrations were initially studied analytically by Facchinetti et al. \cite{facchinetti2004coupling} where they considered a lower order equation for structural vibration and wake vibration (van der Pol equation). Authors considered three different models for coupling wake and structure oscillations through velocity, displacement and inertia terms. The analytical results from these three models were compared with experimental results for various parameters. It was observed that inertia model based on acceleration coupling yielded the best results. A major disadvantage of these models was the ignorance of some terms during derivation (like non-linear variables) and approximation of some parameters based on experiments. Hence, a computational approach applicable to all set of problems becomes necessary.
	\\\\
	Herjfold et al. \cite{herfjord1999assessment} developed a cost effective computational methodology for analyzing the VIV phenomenon on deepwater risers placed in offshore plants so as to test the efficiency of riser design before actually developing it. Authors considered finite number of two dimensional fluid planes along the length of riser on which incompressible Navier-Stokes equations were solved separately. Fractional step algorithm was used using NAVSIM program to solve for velocity and pressure. The values of surface force obtained on fluid simulation were interpolated on surface of structure. Displacement field for solid was computed based on the interpolated surface forces using algorithm of USFOS (Ultimate strength of framed offshore structures). Validation of the developed model was performed by comparing structure response at various sections with the one obtained from experiments. A small amount of under prediction was observed at bottom of the riser. This might have happened due to ignorance of ground effects.
	\\\\
	Zhu and Peng \cite{zhu2009mode} developed a computational model to analyse the effect of heaving motion of foil for the purpose of efficient vortex control. The mathematical formulation of Navier-Stokes equation was performed in XY-coordinate system which was fixed in space and a xy-coordinate system which was fixed with foil. Discretization was performed based on finite difference. In order to incorporate interaction between fluid and structure, an iterative coupling model was used. This model followed some steps wherein initial excitation was provided to the foil and which the deflection (heave) of foil was computed based on the lift force acting on the surface of foil at different sections. They came to a conclusion that partial recovery of energy of flapping could be recovered when leading edge vortex is made to be formed as far as possible from the pitching axis. 
	\\
	Williamson and Govardhan \cite{williamson2004vortex} discussed the overall development in terms of computational and experimental methods to study the VIV phenomenon so as to apply the findings in industrial applications. Review of methods was performed based on their effectiveness to satisfy the Griffin-plot and other parameters. It has been observed that though sufficient computational techniques have been developed to address VIV, much efforts on high Reynolds number VIV with inclusion of turbulence model through Direct Numerical Solution or other methods is the need of the hour to appreciate the real world flows. This might lead us to efficiently develop Industrial products with more confidence. 
	
	\subsection{Aquatic Locomotion and Insect Flight}
	
	Requirements for Enhancing the lift and reducing drag in mechanical systems for locomotion has led the scientific community to find inspiration from aquatic and flying species on Earth. An aerodynamic model was proposed by Weis-Fogh \cite{weis1973quick} through which small insects like wasps were found to enhance lift during their flight. He proposed that insects used clap and fling motion during their flight to optimise performance. Miller and Peskin \cite{miller2005computational} used a ‘target boundary’ version of Immersed Boundary Method to simulate clap and fling in a tiny wasp (called Encarsaria formosa). They performed simulation at various flow conditions (8<=Re<=128) to analyse the lift enhancement mechanism. Displacement and rotation of the wings were specified to understand the interaction between fluid and wing structure. Lift and Drag coefficients were obtained as the outputs which were validated with experimental results. It was found that leading and trailing edge vortices existed for single wing case throughout the cycle whereas only leading edge vortex (LEV) was present in two wing case during initial fling motion. Vortical asymmetry during translation and dominance of leading edge vortex were found to be the reason for lift enhancement. Results obtained for one wing and two wings at various reynolds number showed that clap and fling mechanism was most suitable for insects flying at lower speeds.
	\\
	\\
	Efficient propulsion found in aquatic animals is attributed to the optimal vortex control done by them through movement of body and tail fin. In order to understand the mechanism behind effective propulsion system of fish and derive the optimal parameters for body motion, Triantafyllou \cite{triantafyllou2000hydrodynamics} performed a systematic review of experimental observations and computational methods. It was shown that when an oncoming vortex moves over a foil (which has generated its own vortex), it can either interact constructively to form a large trailing edge vortex or interact destructively to weaken the vortex or form pairs to broaden the wake region. The experiments carried out on oscillating and thrust producing foils showed that there was an optimum St (around 0.25 - 0.35) where maximum propulsive efficiency was noted.
	\\
	\\
	Bhalla et al. \cite{bhalla2013unified} developed a computational Fluid-Structure Interaction model using cartesian based Immersed Boundary Method and applied Adaptive Mesh Refinement to capture thin Boundary Layer at Fluid surface Interface. They validated their model by applying it on a 3D model of black ghost knifefish. The fluid was modelled in an Eulerian framework whereas solid was modelled through Lagrangian framework. Specified motion is provided to the body and force term fc is incorporated in governing equations to account for that whereas a term fb is used to account for deformations of solid body. Reverse karman vortex street is observed in eel locomotion close to trailing edge which suggests the self propelling nature observed in eels. Forward and Backward motion of black ghost knifefish was simulated to understand its ability to swim in the reverse direction.
	\\
	\\
	Recently Thekkethil et al.\cite{thekkethil2018unified} performed a unified hydrodynamics study based on various types of motions (pitch, heave and undulation) through an in-house code based on Level Set Immersed boundary formulation. Apart from solving regular Navier stokes equation, two separate equations based on advection velocity at the interface and level-set function were used to classify the domains. A non dimensional number based on wavelength of undulation was used for unifying various motions. Simulations were performed at several St and propulsive efficiency was derived. The role of reverse Karman street in enhancement of propulsion was described by authors. A sensing mechanism used by predator fish was also proposed after visualising the vortex in the downstream direction. Performance characteristics of self propulsive motion of NACA airfoil (which represents fish cross-section) has been investigated by Thekkethil et al. \cite{thekkethil2020self} using the in-house developed Level Set (LS) based Immersed Interface Method. It was observed that pitching type fishes have higher propulsive efficiency and stability during the initial phases of propulsion. They extended the study to 3D batoid type fish motion \cite{thekkethil2020three} with an added parameter of 3D aspect ratio. Horshoe type vortices with multiple vortex rings were observed for pitching type batoids whereas the propulsive efficiency was found to be high for intermediate wavelength numbers and larger aspect ratios.
	
	\subsection{Hemodynamic Analysis of pulsatile blood flow}
	Cardiovascular ailments have become a major reason for deaths in the modern world mainly due to the sedentary lifestyle and food habits. It was estimated by World Health Organisation (WHO) that around 7.3 million deaths are reported annually around the globe due to coronary artery abnormalities \cite{world2002world}. Many numerical and experimental works have shown there exists a strong correlation between Hemodynamic parameters like Wall Shear Stress(WSS), Oscillatory Shear Index(OSI) , Relative Residence Time (RRT) etc. on the growth and progress of diseases. Cardiac-Surgeons are usually referred to as Cardio-Vascular Fluid mechanicians who brings the blood flow to its normal condition either by inserting a mechanical device which enlarges artery(called stent) or by making some other way for blood flow(Grafting). Since in-vivo measurements are dangerous and very difficult, it becomes utmost build a most accurate computational model which can simulate blood flow through arteries so as to predict potential sites for disease progression, help surgeons in surgery planning and study the causes for disease development. Major challenges for computational development include using patient specific models, using proper rheology of blood, modelling blood constituents and considering proper compliance of arteries through accurate artery material properties.
	\\
	Inclusion of compliance leads us to requirement of coupling fluid and structure interaction. Crosetto et al.\cite{crosetto2011fluid} used ALE (Section 3.1) to calculate the variation of mean velocity and pressure at different section with respect to the cardiac cycle. Patient specific arteries were modelled using ITK Snap software. Simulations were performed through LifeV multi-physics software. A hybrid Boundary condition based on pressure and traction was formulation to obtain realistic conditions. Blood was assumed to have a Newtonian rheology as the arteries considered were of sufficiently larger cross section. It was observed that when the inlet pressure was specified as boundary condition, there was a diastolic backward flow which is not usually observed in fixed wall simulation. This phenomenon was actually reported in human blood circulation cycle which led to closing of aortic valve. It was also shown that WSS was overestimated by rigid wall model. Thus, it can be argued that rigid wall models underestimates the disease growth (as CAD is associated with low and oscillating shear stress zone). 
	\\\\
	Bathe and Kamm \cite{bathe1999fluid} Studied the effects of aortic stenosis severity, asymmetric nature and input pressure conditions on the compressive and tensile stresses, blood flow patterns and plaque rupture possibilities. Computational method developed on Finite element and finite difference scheme was validated with experimental results. Poly Vinyl Alcohol(PVA) - hydrogel was selected as the artery material for experiments. Pressure measurements were obtained through manometric techniques whereas flow rates were measured through ultrasound flow meter. Three dimensional Mooney-Rivlin model was used to model arteries computationally. Fluid domain was solved through finite difference scheme in a curvilinear coordinates. FSI at interfaces was implemented through incremental boundary iteration. Maximum compressive stress was observed near the inner wall at stenosis throat for asymmetric artery whereas point of maximum compression was observed to be in the near post-stenotic region. Turbulence and Non-Newtonian rheology were ignored which have effects on hemodynamic parameters as suggested by Mahalingam et al. \cite{mahalingam2016numerical}.
	\\\\
	Based on some of the fine literature discussed above, it can be observed that there have been minimal efforts to consider all parameters into account to simulate realistic blood flow in compliant arteries. A major facet ignored in currently available studies include the effect of shear induced migration of red blood particles \cite{buradi2019effect} in complaint stenotic arteries in the carotid region.
	
	\subsection{Phonation Study}
	Phonation refers to the production of sound due to fluid induced vibration of the vocal cords in the larynx. Structure of human larynx consists of three different layers \cite{pouderoux1997timing} as shown in figure. Vocal cord paresis occurs due to damage on the upper epithelium and leads to disability in speaking. This problem is usually found during injury or due to old age. In the process of phonation, vocal cord is made to vibrate by the alternate vortex shedding in the wake of the vocal cords. Medialization laryngoplasty is a surgical treatment for vocal cord paresis where surgeon tries to insert an implant made of a small strip which can make up of the damaged mass of epithelium thus enabling normal sound production. This process is iterative in nature and requires a huge experience of surgeon to handle it. Simulation of vocal cords with particular implant can help the surgeon to plan the surgery and remove the iterative procedure.
	\\\\
	In the initial stages, fully coupled FSI approach was used by Ishizaka and Flanagan \cite{ishizaka1972synthesis} where airflow was considered in one dimensional and two lumped masses for vocal cords were used. Mittal et al. \cite{mittal2005immersed} developed a complete two-dimensional FSI model sharp interface immersed boundary formulation. Incompressible NS equations for air discretized through sharp interface immersed boundary method was coupled with cartesian BM formulation performed on structure. Appropriate properties based on experiments were used to model vocal folds. Triangular elements were used to discretize solid and second order finite difference scheme was applied for temporal and spatial discretisation. An equation for displacement eigen mode was formulated and frequency was calculated. Normal modes found through experiments were compared with the eigen modes obtained through simulation. It was observed that second and third eigen modes obtained in simulation were in reversed order from normal modes. Thus, a modification in numerical method to handle solid is needed for accurately solving this issue.
	
	\subsection{Bolus transport through Esophagus}
	Esophagus is a thick multi-layered cylindrical tube which helps to transfer food (called bolus in medical terms) from pharynx to stomach through pulsatile wall motions controlled by neural stimulus. Esophagus is composed of mucosal, Longitudinal muscle (LM) and Circular muscle (CM) layers which shorten and contract simultaneously to produce the desired wall motion \cite{pouderoux1997timing}. It was found that around 5-8 percentage of the population above 50 years of age have swallowing difficulty which is associated with esophageal dysphagia \cite{cook2008diagnostic}. Understanding the role of muscle contraction and shortening in esophageal transport can be achieved through numerical simulations. Quantification of disease pathologies and planning treatment strategies are the next steps in this technology which have to be explored.
	\\\\
	Kou et al. \cite{kou2015fully} developed a computational FSI model to simulate peristaltic wall motions in esophagus using Immersed Boundary approach based on continuous formulation (refer 3.3.2). In this model, bolus was considered as a viscous fluid and esophagus as a thick tube containing fibers at different orientation. This whole setup of fluid inside a solid was then submerged inside another fluid where equations were not solved. Numerical method was validated with analytical solution for tube dilation problem. The Immersed Boundary method developed contains momentum and incompressibility of coupled fluid and structure in an Eulerian framework whereas structural forces are modelled through Lagrangian framework.The three different fibers of esophagus are modelled as springs and beams whose whose stiffness parameter was derived through material properties (Young’s modulus and poisson’s ratio). Finite element formulation of wall structure using triangular elements has been performed based on the shape function calculated through local coordinates. Muscle activation kinematics is used for wall motion through which displacement is specified at the inner layers of Esophagus. Numerical challenges due to a wide range of dimensional quantities in wall diameter and esophagus length were observed due to which some amount of  fluid leakage was observed. An adaptive mesh technique may prevent this issue and keep computational cost in check. Kou et al.\cite{kou2017simulation} validated their numerical pressure contours with clinical manometry measurements.
	
	\section{Conclusions and Future Scope}
	
	\subsection{Conclusions}
	
	A brief overview of different numerical methods used to solve the multi-physics problems of fluid and structure have been discussed with some important application domains where computational methods become very essential. The scope and limitations of the various methods classified based on coupling manner and discretization have been discussed in detail. Methods suitable for each application domain and issues which persist even today with these models have been listed. A list of standard numerical methods, their developers and major application domains which use these methods has been provided in Table~\ref{table_summary} .\\\\
	
	\begin{table*}[]
		\caption{Summary of Literature for various numerical methods and applications}
		\begin{center}
			\label{table_summary}
			\begin{tabular}{c l l}
				& & \\ 
				\hline
				Numerical \\Method & Significant Developers & Application Domains \\
				\hline
				ALE & Hirt et al.\cite{hirt1974arbitrary},Kennedy et al.\cite{kennedy1982theory} & Free surface flows, Sloshing dynamics, vibration of risers,\\ & & Flapping wing aerodynamics. \\
				DSD/SST & Tezduyar et al. \cite{tezduyar1992new} & Modelling patient specific arterial blood flow, Flow through \\ & & Windsock. \\
				IBM & Peskin \cite{peskin1977numerical} & Blood flow through heart valves, Vortex induced vibration, \\ & & Phonation and Esophagal transport. \\
				IIM & Li and Lai \cite{lai2000immersed} & Hydrodynamics of aquatic locomotion, Intra Vena Cava (IVC) \\ & & flow,  Solidification and crystallization study. \\
				CMM & Figueroa et al. \cite{figueroa2006coupled} & Deformable artery flow. \\
				\hline
			\end{tabular}
		\end{center}
	\end{table*}
	
	\subsection{Future Scope}
	There has been very significant progress in the development of computational techniques and application of the developed methods to design useful mechanical devices which can harvest energy, decrease the vibration of deepwater risers, build better infrastructures and help us to understand the way birds, insects and aquatic animals locomote. Recently, there has been a huge advancement in the development of diagnosis techniques, surgical planning in medical domain which have some contribution from Fluid Structure Interaction based computational models. Here we present scope for future work on computational FSI methods and applications in separate subsections below.
	\\
	\subsubsection{Computational FSI Methods}
	\begin{enumerate}
		\item
		Recently Tsui et al. (2013) have used Finite Volume discretisation for solids in FSI problem and validated the numerical method with flow past a circular cylinder with flexible plate attached on its lee side. These methods have an inherent advantage of obtaining better accuracy at interface since solid and fluid has a common coupling scheme at the interface. These methods are not explored completely in the domain of FSI and especially in the domain of cardio-vascular FSI. Using Eulerian framework for solids and discretizing through FVM remains an unanswered question which can have huge implications in accuracy and computational costs.\\
		\item
		It has been observed in most of the literature that FSI model are used to solve problems where fluid motion is assumed to be laminar. Hence, there are no relevant computational model found which use turbulence model for fluid and then applying two way coupling with solid. If we consider the case of pulsatile blood flow in small arteries like those found in cerebrum and in deep veins, a small constriction may cause the flow to become turbulent. Thus turbulence modelling for these kind of biological FSI problems becomes necessary.\\
		\item
		It has been observed in Immersed Boundary method that though second order accuracy is obtained for smooth interfaces, calculations on sharp and especially discontinuous interfaces (usually found on walls with protrusions) cannot achieve this accuracy. Though there are some other methods like Immersed interface method, blob-projection method which can avoid this issue, a core Immersed Boundary method which can obtain translation invariance and second order accuracy needs to be developed.
	\end{enumerate}
	\subsubsection{Computational FSI Applications}
	
	\begin{enumerate}
		\item
		Phonoangiography is a non-invasive diagnosis technique in which arterial murmurs (called bruits) are analysed to detect abnormalities. Though there was substantial research for qualitative phonoangiography \cite{borisyuk2002experimental}, quantitative methods which help to localise the source and quantify the abnormalities have been developed in the recent past through computational techniques \cite{seo2012coupled}. A major assumption made by them was ignorance of fluid induced vibration of structure which can lead to significant change in sound spectrum. They had considered coronary arteries which are well protected by the rib cage which may substantially reduce the resonance peak created by structure vibration. If we consider the case of carotid arteries, this assumption might not hold true as it is located very close to the epidermal surface where measurements are made. Thus, analysis of bruits on carotid arteries with Fluid Structure model coupled with acoustics can take us close to device a Guided User Interface(GUI) which can help medical practitioners to quantify abnormalities quantitatively.\\
		\item
		Considering Shear Induced Migration (SIM) of Red Blood Cells (RBC) in small complaint arteries and calculating Hemodynamic parameters to predict growth of diseases can help surgeons in medical planning. This problems requires one to couple FSI model for blood-artery with the mixture model which governs transport of RBC to obtain final solution. This becomes very important in small arteries where SIM is enhanced and may change the platelet activation mechanisms which are the main reasons for plaque deposits.\\
		\item
		Development of new energy harvesting devices like tapered vented cylinders as vortex generators need a very efficient FSI model which can handle geometric complexities. It was proposed by Kumar et al. \cite{kumar2016energy} that usage of such systems can improve energy harvesting quantitatively but FSI model was not incorporated by authors due to geometric complexity.\\
		\item
		Performance of new underwater vehicles operating on the strategy of aquatic locomotion needs to be verified before actual design and manufacturing. These include the use of flexible plate on the lee side of cylinder as a fin for these vehicles in order to reduce drag. Though simulation has been performed on part scales, a full fledged simulation using FSI models is very essential for moving ahead with next procedures.\\
	\end{enumerate}

	\bibliographystyle{unsrt}	
	\bibliography{FSIreview}
	
\end{document}